\documentclass{article}
\usepackage{amssymb}
\usepackage{amsmath}

\begin{document}
\centerline{\Large \bf The decomposition method and Maple procedure}
\medskip
\centerline{\Large \bf for finding first integrals of nonlinear
PDEs}
\medskip
\centerline{\Large \bf of any order with any number of independent
variables}
\vskip 2cm \centerline{\sc Yu. N. Kosovtsov}
\medskip
\centerline{Lviv Radio Engineering Research Institute, Ukraine}
\centerline{email: {\tt kosovtsov@escort.lviv.net}} \vskip 1cm
\begin{abstract}
In present paper we propose seemingly new method for finding
solutions of some types of nonlinear PDEs in closed form. The method
is based on decomposition of nonlinear operators on sequence of
operators of lower orders. It is shown that decomposition process
can be done by iterative procedure(s), each step of which is reduced
to solution of some auxiliary PDEs system(s) for one dependent
variable. Moreover, we find on this way the explicit expression of
the first-order PDE(s) for first integral of decomposable initial
PDE. Remarkably that this first-order PDE is linear if initial PDE
is linear in its highest derivatives.

The developed method is implemented in Maple procedure, which can
really solve many of different order PDEs with different number of
independent variables. Examples of PDEs with calculated their
general solutions demonstrate a potential of the method for
automatic solving of nonlinear PDEs.
\end{abstract}

\section{Introduction}

Nonlinear partial differential equations (PDEs) play very important
role in many fields of mathematics, physics, chemistry, and biology,
and numerous applications. If for nonlinear ordinary differential
equations (ODEs) one can observe incontestable progress in their
automatic solving, the situation for nonlinear PDEs seems as nearly
hopeless one.

Despite the fact that various methods for solving nonlinear PDEs
have been developed in 19-20 centuries as the suitable groups of
transformations, such as point or contact transformations,
differential substitutions, and Backlund transformations etc., the
most powerful method for explicit integration of \emph{second-order}
nonlinear PDEs in \emph{two independent variables} remains the
method of Darboux \cite{Darb1}-\cite {Fors}. The original "Darboux
method" (as already Darboux stated in \cite{Darb1}) is extendable in
principle to equations of all orders in an arbitrary number of
independent variables, even to systems of equations; however, in
\cite{Darb1}-\cite{Darb2} and subsequent papers by many authors, the
detailed calculations were performed only for a single second-order
equation with one dependent and two independent variables.

The Darboux method was refined in recent years into more precise and
efficient (although not completely algorithmic) form \cite
{Juras1}-\cite {Zhiber2} and references therein. Nevertheless this
approaches suffer from high complexity and necessitate to use some
tricks.

There were some partially successful attempts to extend modern
variants of the Darboux method based on Laplace cascade method on
higher-order PDEs and PDEs in the space of more than two independent
variables \cite{Roux}-\cite {Tsarev2} but they suffer from high
complexity too.

There is an original approach to the problem, based on the special
type of local change of variables which leads to the order reduction
of initial PDE, proposed in \cite{Boyko}, which is suitable for high
dimensions problems but of very special class though.

In present paper we propose seemingly new method for finding
solutions of some types of nonlinear PDEs in closed form. The method
is based on decomposition of nonlinear operators on sequence of
operators of lower orders. It is shown that decomposition process
can be done by iterative procedure(s), each step of which is reduced
to solution of some auxiliary PDEs system(s) for one dependent
variable. Moreover, we find on this way the explicit expression of
the first-order PDE(s) for first integral of decomposable initial
PDE. Remarkably that this first-order PDE is linear if initial PDE
is linear in its highest derivatives.

The developed method is implemented in Maple procedure, which can
really solve many of different order PDEs with different number of
independent variables. Examples of PDEs with calculated their
general solutions demonstrate a potential of the method for
automatic solving of nonlinear PDEs.

\section{Bases of the method}
\subsection{Decomposable PDEs}
The simplest second-order non-linear
PDE for $w=w(t,x)$
\begin{equation}
\frac{\partial^2 w}{\partial t
\partial x}=\frac{1}{w}\frac{\partial w}{\partial t}\frac{\partial w}{\partial x}
\label{P}
\end{equation}
can be easily transformed to the following decomposed form
\begin{equation}
\frac{\partial }{\partial t} \ln(\frac{1}{w}\frac{\partial
w}{\partial x})=0\,, \label{P1}
\end{equation}
from which we can without difficulty obtain the general solution to
PDE (\ref{P}) in two steps. First step gives us

\begin{equation}
\frac{1}{w}\frac{\partial w}{\partial x}=\frac{d \ln(G(x))}{dx}\,,
\label{AP1}
\end{equation}
where $G(x)$ is an arbitrary function. And then, solving the
equation (\ref{AP1}) on the second step, we obtain
\begin{equation}
w(t,x)=F(t)G(x)\,, \label{AP2}
\end{equation}
where $F(t)$ is one more arbitrary function.

The main observations on analyzing the grounds of solvability of the
PDE (\ref{P}) \emph{by the above} method are that

1. The PDE (\ref{P}) is \emph{"decomposable"}, i.e., it can be
represented as a composition of successive differential operators of
type (\ref{DD}) (not necessarily linear). It is clear that such type
of decomposition can be done for some PDEs of any order and with any
number of independent variables in the following manner
\begin{align}
D_1(w)&=u_1\,,\notag \\ D_2(u_1)&=u_2\,,\notag
\\ \dots&\dots\,, \label{DD} \\ D_n(u_{n-1})&=0\,,\notag
\end{align}
where $\vec{x}=(x_1,\dots , x_m)$, $w=w(\vec{x})$,
$u_i=u_i(\vec{x})$ and
\begin{equation}
D_i(u)=V_i(\vec{x},u,\frac{\partial u}{\partial
x_1},\dots,\frac{\partial u}{\partial x_m})\,.\notag
\end{equation}
Assuming that $V_i$ are arbitrary functions, and eliminating $u_i$
by successive substitutions in system (\ref{DD}), we get a family of
PDEs for $w$ of $n$th order
\begin{equation}
D_n(D_{n-1}(\dots D_1(w)\dots))=0\,.\label{PDE}
\end{equation}
which are "decomposable" and \emph{in principle} their solutions
\emph{general} or \emph{particular} can be obtained by integration
of split system (\ref{DD}). The PDE (\ref{PDE}) is nonlinear if at
least one of the operators $D_i$ is nonlinear. Not all PDEs admit
such representation. And in positive cases such representation is
not unique in general.

Note that as a matter of fact $D_i$ need not be the first-order
differential operators. So the composition procedure for $n$th order
PDE, when $n>2$ can be as follows
\begin{align}
D_1^{n_1}(w)&=u\,,\notag \\ D_2^{n_2}(u)&=0\,,\label{DD2}
\end{align}
where $n_1$, $n_2$ are integers and $n_1+n_2=n$, $w=w(\vec{x})$,
$u=u(\vec{x})$, and ($k\leq j$)
\begin{equation}
D_i^j(u)=V_i(\vec{x},u,\frac{\partial u}{\partial
x_1},\dots,\frac{\partial^k u}{\partial x_1^{k_1}\dots\partial
x_m^{k_m}|_{k_1+\dots +k_m=k\leq j}},\dots,\frac{\partial^j
u}{\partial x_m^j})\,.\notag
\end{equation}

The late representation allows us to carry out the PDE`s
decomposition or order reduction \emph{gradually} bit by bit.

We have to stress here that in general representations (\ref{DD})
and (\ref{DD2}) may have different meaning. For example, some PDEs
do not admit representation (\ref{DD}) but permit the form
(\ref{DD2}) with both solvable DEs.

2. Each step of the solving process for decomposed PDE is faced with
the necessity to solve differential equation $ D_i(u_{i-1})=u_i$ (or
$ D_i^j(u_{i-1})=u_i$), so all such DEs must be solvable. Note that
only first step $D_n(u_{n-1})=0$ is \emph{free from arbitrary
functions}.

So one of the PDEs solving strategies may be as follows. First of
all we try to decompose given PDE. In order to do so we have to
solve corresponding auxiliary nonlinear PDE system for unknown
functions $V_i$, it is sufficient to find a particular solution
here. And, if it is successful, then, deciding between the variants,
try to solve each arising DE from the chain (\ref{DD}). Main
obstacle here, beginning at the second step is just mentioned
necessity to solve DEs with arbitrary functions. There are
sufficiently narrow circle of solvable (in sense of the general
solutions) DEs with an arbitrary function as a parameter.

Another (classification) approach can be based on the usage of only
solvable DEs. That is, we can form a composition of successive
\emph{solvable} differential operators and as a result obtain a
families of solvable PDEs. Such a way leads to extensive nontrivial
families for different types of nonlinear PDEs which general
solutions can be expressed in closed form. But on this way we
encounter a difficulty to circumscribe such families integrally and
are forced to consider particular subfamilies. Nevertheless it
yields extensive field of PDEs for methods testing \cite {Kos}.

\subsection{Decomposition algorithm for decomposable PDEs}

For $n$th order PDE, when $n>2$ there are some slightly different
approaches which are dictated by goals of the problem. If the goal
is to \emph{decompose} given nonlinear operator then we have to use
the scheme (\ref{DD2}) with $n_1=1$, $n_2=n-1$. And conversely we
have to use the scheme (\ref{DD2}) with $n_1=n-1$, $n_2=1$ if the
goal is to \emph{solve} given PDE. The last procedure in some
features resembles the well-known technics of reducing ODEs order,
e.g., by first integral method. Of course, it is possible to use
intermediate cases.

All above cases can be treated by the same way as we consider below
but each of them leads to auxiliary PDEs systems of different order,
viz $n_2+1$, with corresponding calculation complexity.

In sequel we will consider for shortness only the case with
$n_1=n-1$, $n_2=1$, as more practical for PDEs solving.

Let us consider the decomposition of type (\ref{DD2}) with
$D_1^{n-1}(w)$ as a solution of the following equation with respect
of $u=u(\vec{x})$
\begin{equation}
J(u,\vec{x},w,\frac{\partial w}{\partial x_1},\dots,\frac{\partial^k
w}{\partial x_1^{k_1}\dots\partial x_m^{k_m}|_{k_1+\dots +k_m=k\leq
n-1}},\dots,\frac{\partial^{n-1} w}{\partial x_m^{n-1}})=0
\label{DJ1}
\end{equation}
and
\begin{equation}
D_2(u)=V(\vec{x},u,\frac{\partial u}{\partial
x_1},\dots,\frac{\partial u}{\partial x_m})\,.\label{DJ2}
\end{equation}
If substitute $u=D_1^{n-1}(w)$ into (\ref{DJ2}) we obtain
decomposable n-th order PDE
\begin{equation}
V(\vec{x},U_0,U_{x_1},\dots,U_{x_m})=0\,,\label{DJ3}
\end{equation}
where (we use below the following notation $w=W_0$ and
$\frac{\partial^k w}{\partial x_1^{k_1}\dots\partial
x_m^{k_m}}=W_{k_1,\dots,k_m}$)
\begin{equation}
D_1^{n-1}(w)=U_0\,,\label{DJ32}
\end{equation}
\begin{equation}
-\frac{\frac{\partial J}{\partial x_i}+\sum \frac{\partial
J}{\partial W_{k_1,\dots,k_m}}W_{k^*_1,\dots,k^*_m}}{\frac{\partial
J}{\partial u}}=U_{x_i}\qquad (i=1,\dots ,m)\,,\label{DJ32}
\end{equation}
where $k^*_j=k_j+1$ if $j=i$ and $k^*_j=k_j$ otherwise, and it is
supposed that differentiations in sum are carried out on all indexed
$W$`s which are involved in $J$.

Here we can introduce $U_0$ and $U_{x_1},\dots,U_{x_m}$ as new
independent variables if express $m$ variables from the set
$\{W_{k_1,\dots,k_m}\}$ with ${k_1+\dots +k_m= n}$ using linear
system (\ref{DJ32}).

Assuming that given PDE of order $n$
\begin{equation}
F(\vec{x},w,\frac{\partial w}{\partial x_1},\dots,\frac{\partial^k
w}{\partial x_1^{k_1}\dots\partial x_m^{k_m}|_{k_1+\dots +k_m=k\leq
n}},\dots,\frac{\partial^n w}{\partial x_m^n})=0 \label{DJ4}
\end{equation}
is decomposable, we receive, that after substitution of the new
variables, \emph{left-hand side} of given PDE must turn into
(\ref{DJ3}) with some $V$.

\emph{Left-hand side} of given PDE expressed in new variables is the
first-order differential expression with respect to
\[J(U_0,\vec{x},W_0,W_{1,0,\dots,0},\dots,W_{k_1,\dots,k_m}|_{k_1+\dots
+k_m=k\leq n-1},\dots,W_{0,0,\dots,n-1})\] and must not depend on
all indexed $W$`s, that is derivatives of $F$ expressed in new
variables with respect to all indexed $W$`s are equal to zero.
Sequence of such derivatives of $F$ equated to zero form a
\emph{second-order} PDE system for $J$. So a solution (particular as
well) the PDE system gives possible expression of differential
operator $D_1^{n-1}(w)$ through (\ref{DJ1}) and differential
operator $D_2(u)$ by substituting the solution of $J$ into
\emph{left-hand side} of given PDE expressed in new variables.

Of course, there are problems where a operator decomposition is
required only. But in most cases obtained decomposition is intended
for finding solutions for given PDE. If in obtained decomposition
the corresponding PDE $D_2(u)=0$ is solvable, then substitution of
obtained $u$ into $J$ expressed in original variables gives us
\emph{a first integral} (see its definition in the next subsection)
of given PDE. It is easy to see that for decomposable PDEs the first
integral is a differential equation, so we can try to solve it or to
find a first integral for this new DE (or decompose it) by the
scheme described above until we come to the first-order DE.

Remarkably that in the approach under consideration the finding of
first integrals can be done more directly and effectively.

\subsection{Differential equation for first integral of decomposable PDEs}

The first integral $I$ of the PDE is an expression, involving
\emph{one} arbitrary function, which is equivalent in some sense to
the given PDE. The first integral vanishes on the set of solutions
of given PDE. And (in accordance with \cite {Fors}) all differential
consequences of the equation $I=0$ coincide with respective
differential consequences of given PDE (e.g., elimination of the
arbitrary function leads to the given PDE).

Our goal here is to find PDE for first integral of a decomposable
PDE. To do so we first of all have to take into account that
$u(\vec{x})$ is the solution of the corresponding PDE
\begin{equation}
V(\vec{x},u,\frac{\partial u}{\partial x_1},\dots,\frac{\partial
u}{\partial x_m})=0\,,\notag
\end{equation}
so $u(\vec{x})$ depends only on $\vec{x}$ but in no way on indexed
$W$`s. Secondly, the dependent variable in this case, namely
\[J(u(\vec{x}),\vec{x},W_0,W_{1,0,\dots,0},\dots,W_{k_1,\dots,k_m}|_{k_1+\dots
+k_m=k\leq n-1},\dots,W_{0,0,\dots,n-1})\] of given PDE (\ref{DJ4})
expressed in new variables do not to depend on
$U_{x_1},\dots,U_{x_m}$ and is \emph{a first integral} of given PDE.

If now consider $u(\vec{x})$ as an unknown function, we can denote
the first integral as
\begin{align}
&I(\vec{x},W_0,W_{1,0,\dots,0},\dots,W_{k_1,\dots,k_m}|_{k_1+\dots
+k_m=k\leq n-1},\dots,W_{0,0,\dots,n-1})=\notag
\\&J(u(\vec{x}),\vec{x},W_0,W_{1,0,\dots,0},\dots,W_{k_1,\dots,k_m}|_{k_1+\dots
+k_m=k\leq n-1},\dots,W_{0,0,\dots,n-1})\notag
\end{align}
and instead of (\ref{DJ32}) in the form
\begin{equation}
\frac{\partial J}{\partial x_i}+\sum \frac{\partial J}{\partial
W_{k_1,\dots,k_m}}W_{k^*_1,\dots,k^*_m}=-U_{x_i}\frac{\partial
J}{\partial u}\qquad (i=1,\dots ,m)\notag
\end{equation}
we arrive to the following system
\begin{equation}
\frac{\partial I}{\partial x_i}+\sum \frac{\partial I}{\partial
W_{k_1,\dots,k_m}}W_{k^*_1,\dots,k^*_m}=0\qquad (i=1,\dots
,m)\,.\label{I1}
\end{equation}

If express $m$ variables from the set $\{W_{k_1,\dots,k_m}\}$ with
${k_1+\dots +k_m= n}$ (at least one of which is actual for given PDE
- note that there are some variants here as a rule, so we can obtain
\emph{some} consistent PDEs on this stage) using linear system
(\ref{I1}) and substitute them into given PDE (\ref{DJ4}) we receive
a \emph{first-order} (even \emph{linear} if PDE (\ref{DJ4}) is
linear in its highest derivatives) PDE with respect to first
integral $I$. And it remains only to solve this PDE(s) to find a
first integral of given PDE.

Note, given PDE is decomposable iff exists a solution of such
first-order PDE(s).

\section{Examples}

To facilitate necessary calculations in the process of finding first
integrals I have implemented above described method in
\emph{prototype} of \emph{Maple} procedure \emph{reduce\_PDE\_order}
(see Appendix). The input data of the procedure are given PDE of any
order and dependent variable of the PDE with any number of
independent variables. The procedure tries to find first integral(s)
of the input linear or nonlinear PDE.

The \emph{Maple} built-in procedure \emph{pdsolve} is used inside my
procedure to solve \emph{the first-order} PDE for first integral. As
different \emph{Maple} versions have different PDE solving abilities
so the output depends on \emph{Maple} version. In the following
examples I refer to \emph{Maple} 11.

The procedure \emph{reduce\_PDE\_order} is able to find first
integrals for many known and unknown linear and nonlinear PDEs. Here
we give examples of PDEs for which it is possible to find finally
their \emph{general} solutions. More examples one can find in
collection of solvable nonlinear PDEs \cite {Kos}.

\subsection{Second-order PDE with two independent variables}

For PDE ($w=w(t,x)$)
\begin{equation}
\displaystyle \frac{\partial^2 w}{\partial t\partial
x}-\frac{a}{w}\left(\frac{\partial w}{\partial
x}\right)^{\!2}-\left(\frac{1}{w}\frac{\partial w}{\partial
t}+b+\frac{c}{w}\right)\frac{\partial w}{\partial x}
-\frac{c}{2aw}\frac{\partial w}{\partial
t}-kw-\frac{bc}{2a}-\frac{c^2}{4aw} = 0\label{EQ1}
\end{equation}
with $a\neq0$ and $4ak-b^2\neq0$ the procedure
\emph{reduce\_PDE\_order} outputs the following first integral
\[I=F1\left[x,\frac{t\sqrt{4ak-b^2}-2\arctan
\displaystyle\left(\frac{c+2a\frac{\partial w}{\partial x}+b
w}{w\sqrt{4ak-b^2}}\right)}{\sqrt{4ak-b^2}}\right]\] with arbitrary
function $F1$.

The ODE $I=0$ can be solved and one obtains (after some hand
simplifications and edition) the following \emph{general} solution
to (\ref{EQ1})
\begin{align}
\displaystyle &w(t,x) =\notag\\
&-\frac{c}{2a}\{\int\exp\left[\frac{1}{2a}\int\frac{\exp(t\sqrt{b^2-4ak})F(x)(b+\sqrt{b^2-4ak})-\sqrt{b^2-4ak}+b}{1+\exp(t\sqrt{b^2-4ak})F(x)}dx\right]dx
+\notag\\\notag\\
&G(t)\}
\exp\left[-\frac{1}{2a}\int\frac{\exp(t\sqrt{b^2-4ak})F(x)(b+\sqrt{b^2-4ak})-\sqrt{b^2-4ak}+b}{1+\exp(t\sqrt{b^2-4ak})F(x)}dx\right]\,,\notag
\end{align}
where $F(x)$ and $G(t)$ are arbitrary functions.

\subsection{Second-order PDE with four independent variables}

For PDE
\begin{align}
\displaystyle &A_1\frac{\partial^2 w}{\partial x_1
\partial x_4}+A_2\frac{\partial^2 w}{\partial x_2
\partial x_4}+A_3\frac{\partial^2 w}{\partial x_3
\partial x_4}+C_0+B_1\frac{\partial w}{\partial x_4}+\notag\\&C_1(A_1\frac{\partial w}{\partial x_1}+A_2\frac{\partial w}{\partial x_2}+A_3\frac{\partial w}{\partial x_3}+B_1w+B_0)+\notag\\&C_2(A_1\frac{\partial w}{\partial x_1}+A_2\frac{\partial w}{\partial x_2}+A_3\frac{\partial w}{\partial
x_3}+B_1w+B_0)^2=0\,,\label{EQ2}
\end{align}
where $w=w(x_1,x_2,x_3,x_4)$ and $A_i,B_i,C_i$ are constants, the
procedure \emph{reduce\_PDE\_order} outputs the following first
integral
\begin{equation}
I=F1\left[x_1,x_2,x_3,x_4+\frac{2\arctan\left(\displaystyle
\frac{2C_2(A_1\frac{\partial w}{\partial x_1}+A_2\frac{\partial
w}{\partial x_2}+A_3\frac{\partial w}{\partial
x_3}+B_1w+B_0)+C_1}{\sqrt{4C_0C_2-C_1^2}}\right)}{\sqrt{4C_0C_2-C_1^2}}\right]\notag
\end{equation}
with arbitrary function $F1$.

The PDE $I=0$ can be solved and one obtains the following
\emph{general} solution to (\ref{EQ2})
\begin{align}
&w(x_1,x_2,x_3,x_4) =\notag \\&
-\frac{1}{2A_1C_2}\exp(-\frac{B_1x_1}{A_1})\int_c^{x_1}\exp(\frac{B_1\xi}{A_1})(2B_0C_2+C_1+\tan[\frac{1}{2}x_4\sqrt{4C_0C_2-C_1^2}\notag
\\ \notag \\&+G(\xi,(
A_2\xi+A_1 x_2-A_2
x_1),(A_3\xi+A_1x_3-A_3x_1))]\sqrt{4C_0C_2-C_1^2})d\xi\notag
\\ \notag \\&
+\exp(-\frac{B_1 x_1}{A_1})F[(A_1x_2-A_2x_1),(A_1x_3-A_3x_1),x_4]\,,
\notag
\end{align}
where $F(t_1,t_2,t_3)$ and $G(t_1,t_2,t_3)$ are arbitrary functions,
$c$ is arbitrary constant.

\subsection{Third order PDE with two independent variables}

For PDE ($w=w(t,x)$)
\begin{equation}
\displaystyle w^2\frac{\partial^3 w}{\partial t\partial
x^2}-2w\frac{\partial^2 w}{\partial t\partial x}\frac{\partial
w}{\partial x}+2\frac{\partial w}{\partial t}\left(\frac{\partial
w}{\partial x}\right)^2-w\frac{\partial w}{\partial
t}\frac{\partial^2 w}{\partial x^2}-a w^3 = 0\label{EQ3}
\end{equation}
the procedure \emph{reduce\_PDE\_order} outputs the following first
integrals
\begin{equation}
I_1=F1\left[t,\frac{1}{w^2}(w\frac{\partial^2 w}{\partial t\partial
x}-\frac{\partial w}{\partial t}\frac{\partial w}{\partial x}-ax
w^2),\frac{1}{w^2}\left[ax^2w^2+2w(\frac{\partial w}{\partial
t}-x\frac{\partial^2 w}{\partial t\partial x})+2x\frac{\partial
w}{\partial t}\frac{\partial w}{\partial x}\right]\right]\notag
\end{equation}
and
\begin{equation}
I_2=F1\left[x,\frac{1}{w^2}\left[w\frac{\partial^2 w}{\partial
x^2}-at w^2-\left(\frac{\partial w}{\partial
x}\right)^2\right]\right]\notag
\end{equation}
with arbitrary function $F1$.

We can form some PDEs from $I_1$ and to solve them we can repeat the
process of order reduction with the procedure
\emph{reduce\_PDE\_order}. The ODE $I_2=0$ can be solved directly
and one obtains in any way the following \emph{general} solution to
(\ref{EQ3})
\begin{equation}
\displaystyle w(t,x) = F(t)\exp \left(\displaystyle\frac{atx^2}{2}-
xH(t)+x\int G(x)dx-\int x G(x)dx\right)\,,\notag
\end{equation}
where $F(t)$, $H(t)$ and $G(x)$ are arbitrary functions.

\subsection{Fourth order PDE with two independent variables}

For PDE ($w=w(t,x)$)
\begin{align}
\displaystyle w^3&\frac{\partial^4 w}{\partial t^2\partial
x^2}-2w^2\left(\frac{\partial^3 w}{\partial t^2\partial
x}\frac{\partial w}{\partial x}+\frac{\partial^3 w}{\partial
t\partial x^2}\frac{\partial w}{\partial
t}\right)-2\left(w\frac{\partial^2 w}{\partial t\partial
x}-2\frac{\partial w}{\partial t}\frac{\partial w}{\partial
x}\right)^2+\notag\\\notag\\&2\left[w\frac{\partial^2 w}{\partial
x^2}+\left(\frac{\partial w}{\partial
x}\right)^2\right]\left(\frac{\partial w}{\partial
t}\right)^2-w\frac{\partial^2 w}{\partial
t^2}\left[w\frac{\partial^2 w}{\partial x^2}-2\left(\frac{\partial
w}{\partial x}\right)^2\right]= 0\label{EQ4}
\end{align}
the procedure \emph{reduce\_PDE\_order} outputs the following first
integrals
\begin{align}
&I_1=F1(t,\frac {1}{w^3}\left[w^2\frac{\partial^3 w}{\partial
t^2\partial x}-2w\frac{\partial w}{\partial t}\frac{\partial^2
w}{\partial t\partial x}-w\frac{\partial^2 w}{\partial
t^2}\frac{\partial w}{\partial x}+2\frac{\partial w}{\partial
x}\left(\frac{\partial w}{\partial
t}\right)^2\right],\notag\\\notag\\&\frac
{1}{w^3}\left[\left(\frac{\partial^2 w}{\partial
t^2}-x\frac{\partial^3 w}{\partial t^2\partial
x}\right)w^2+\left[\left(2\frac{\partial w}{\partial
t}\frac{\partial^2 w}{\partial t\partial x}+\frac{\partial^2
w}{\partial t^2}\frac{\partial w}{\partial
x}\right)x-\left(\frac{\partial w}{\partial
t}\right)^2\right]w-2x\frac{\partial w}{\partial
x}\left(\frac{\partial w}{\partial t}\right)^2\right])\notag
\end{align}
and
\begin{align}
&I_2=F1(x,\frac {1}{w^3}\left[w^2\frac{\partial^3 w}{\partial
t\partial x^2}-2w\frac{\partial w}{\partial x}\frac{\partial^2
w}{\partial t\partial x}-\frac{\partial w}{\partial
t}\frac{\partial^2 w}{\partial x^2}w+2\frac{\partial w}{\partial
t}\left(\frac{\partial w}{\partial
x}\right)^2\right],\notag\\\notag\\&\frac
{1}{w^3}\left[\left(\frac{\partial^2 w}{\partial
x^2}-t\frac{\partial^3 w}{\partial t\partial
x^2}\right)w^2+\left[\left(2\frac{\partial w}{\partial
x}\frac{\partial^2 w}{\partial t\partial x}+\frac{\partial^2
w}{\partial x^2}\frac{\partial w}{\partial
t}\right)t-\left(\frac{\partial w}{\partial
x}\right)^2\right]w-2t\frac{\partial w}{\partial
t}\left(\frac{\partial w}{\partial x}\right)^2\right])\notag
\end{align}
with arbitrary function $F1$.

The wealth of first integrals here allows us to operate with them in
many different ways. Apart from aforesaid subsequent order reduction
we can, for example, from
\[\frac
{1}{w^3}\left[w^2\frac{\partial^3 w}{\partial t\partial
x^2}-2w\frac{\partial w}{\partial x}\frac{\partial^2 w}{\partial
t\partial x}-\frac{\partial w}{\partial t}\frac{\partial^2
w}{\partial x^2}w+2\frac{\partial w}{\partial t}\left(\frac{\partial
w}{\partial x}\right)^2\right]=F(x)\] and
\begin{align}
\frac {1}{w^3}[\left(\frac{\partial^2 w}{\partial
x^2}-t\frac{\partial^3 w}{\partial t\partial
x^2}\right)w^2+&\left[\left(2\frac{\partial w}{\partial
x}\frac{\partial^2 w}{\partial t\partial x}+\frac{\partial^2
w}{\partial x^2}\frac{\partial w}{\partial
t}\right)t-\left(\frac{\partial w}{\partial
x}\right)^2\right]w-\notag\\&2t\frac{\partial w}{\partial
t}\left(\frac{\partial w}{\partial x}\right)^2]=G(x)\,,\notag
\end{align}
where $F(x)$ and $G(x)$ are arbitrary functions, algebraically
eliminate mixed derivative and obtain the following ODE

\[w\frac{\partial^2 w}{\partial x^2}-\left(\frac{\partial w}{\partial
x}\right)^2+\left[t F(x)-G(x)\right]w^2=0\,,\] which gives the
\emph{general} solution to (\ref{EQ4})
\begin{align}
w(t,x) = H(t) &\exp \left [t\int xF(x)\,dx-tx\int F(x)\,dx +\right.
\notag
\\& \left. x\int G(x)\,dx-\int xG(x)\,dx+xK(t)\right]\,,\notag
\end{align}
where $F(x)$, $H(t)$, $G(x)$ and $K(t)$ are arbitrary functions.

\section{Conclusion}

The method have considered above is efficient enough for solving
decomposable PDEs of relatively high order with many independent
variables. The main limitation here is concerned with abilities to
solve corresponding auxiliary first-order PDEs for first integrals.

An adaptability of the method to PDEs which are not decomposable but
which general solutions can be expressed in closed form remains
unsolved yet. But it can be shown on examples that there are some
ways to extend the method for some types of such PDEs. These
approaches deserve further thorough study in another publication.

\section{Appendix. \\\emph{Maple} procedure \emph{reduce\_PDE\_order}}
reduce\_PDE\_order:=proc(pde,unk) \\
local
B,W,N,NN,ARG,acargs,i,M,pde0,DN,IND,IND2,IND3,IND4,ARGS,SUB,SUB0,
Z0,Bargs,EQS,XXX,WW,BB,PP,pdeI,IV,s,AN;

 option `Copyright (c) 2006-2007 by Yuri N. Kosovtsov. All rights reserved.`;
 \\N:=PDETools[difforder](op(1,[selectremove(has,indets(pde,function),unk)]));
 \\NN:=op(1,[selectremove(has,op(1,[selectremove(has,indets(pde,function),unk)]),diff)]);
 ARG:=[op(unk)];
 \\acargs:=\{\};
 \\for i from 1 to nops(ARG) do
 \\if PDETools[difforder](NN,op(i,ARG))=0 then else acargs:=acargs union \{op(i,ARG)\} fi; od;
 \\acargs:=convert(acargs,list);
 \\M:=op(0,unk)(op(acargs));
 \\if type(pde,equation)=true then
 \\pde0:=lhs(subs(unk=M,pde))-rhs(subs(unk=M,pde)) else pde0:=subs(unk=M,pde) fi;
 \\DN:=[seq(seq(i,i=1..nops(acargs)),j=1..N)];
 \\IND:=seq(op(combinat[choose](DN,i)),i=1..N);
 \\IND2:=seq(op(combinat[choose](DN,i)),i=1..N-2);
 \\IND3:=op(combinat[choose](DN,N-1));
 \\IND4:=op(combinat[choose](DN,N));
 \\ARGS:=op(unk),M,seq(convert(D[op(op(i,[IND2]))](op(0,unk))\\(op(acargs)),diff),i=1..nops([IND2]));
 \\SUB:=\{M=W[0],seq(convert(D[op(op(i,[IND]))](op(0,unk))\\(op(acargs)),diff)=W[op(op(i,[IND]))],i=1..nops([IND]))\};
 \\SUB0:=\{W[0]=op(0,unk)(op(ARG)),\\seq(W[op(op(i,[IND]))]=subs(M=op(0,unk)(op(ARG)),\\convert(D[op(op(i,[IND]))](op(0,unk))(op(acargs)),diff)),i=1..nops([IND]))\};
 \\Z0:=B(ARGS,seq(convert(D[op(op(i,[IND3]))](op(0,unk))(op(acargs)),diff),\\i=1..nops([IND3])));
 \\Bargs:=op(indets(subs(SUB,Z0),name));
 \\EQS:=convert(subs(SUB,\{seq(diff(Z0,op(i,acargs))=0,i=1..nops(acargs))\}),diff);
 \\XXX:=\{seq(W[op(op(i,[IND4]))],i=1..nops([IND4]))\};
 \\WW:=select(type,indets(subs(SUB,pde0)), 'name') intersect\\ \{seq(W[op(op(i,[IND4]))],i=1..nops([IND4]))\};
 \\BB:=select(has,combinat[choose](XXX, nops(acargs)),WW);
 \\PP:=\{\};
 \\pdeI:=\{seq(\{subs(subs(solve(EQS,op(i,BB)),subs(SUB,pde0)))\},i=1..nops(BB))\};
 \\IV:=\{seq(W[op(op(i,[IND4]))],i=1..nops([IND4]))\};
 \\for s from 1 to nops(pdeI) do
 \\try
 \\AN:=pdsolve(op(s,pdeI),\{B\},ivars=IV);
 \\for i from 1 to nops(AN) do
 \\if op(0,lhs(op(i,AN)))=B then
 \\PP:=PP union \{rhs(op(i,AN))\}
 \\fi;
 \\od;
 \\catch:
 \\end try;
 \\od;
 \\PP:=subs(SUB0,PP);
 \\RETURN(PP);
 \\end proc:

\vskip 1cm \textbf{Calling Sequence}:
\emph{reduce\_PDE\_order}(\textbf{PDE},\,\textbf{$f(\vec{x})$});
\medskip

\textbf{PDE}\,\,-\,\,partial differential equation;

\textbf{$f(\vec{x})$}\,\,-\,\,indeterminate function with its
arguments.

\end{document}